\documentclass[journal,twoside,web]{ieeecolor}
\usepackage{generic}
\usepackage{cite}
\usepackage{amsmath,amssymb,amsfonts}
\usepackage{algorithmic}
\usepackage{graphicx}
\usepackage{subcaption}
\usepackage{textcomp}
\def\BibTeX{{\rm B\kern-.05em{\sc i\kern-.025em b}\kern-.08em
    T\kern-.1667em\lower.7ex\hbox{E}\kern-.125emX}}
\begin{document}
\title{Variability Analysis in a 3-D Multi-Granular Hf$_x$Zr$_{1-x}$O$_2$ Ferroelectric Capacitor }
\author{Nilesh Pandey, Karishma Qureshi, and Yogesh Singh Chauhan
	\thanks{This work was partially supported by the Swarna Jayanti Fellowship (Grant No. – DST/SJF/ETA-02/2017-18) and FIST Scheme (Grant No. – SR/FST/ETII-072/2016) of the Department of Science and Technology.
		N. Pandey, K. Qureshi and Y. S. Chauhan are with the Nanolab, Department of Electrical Engineering, IIT Kanpur, Kanpur 208016, India (e-mail: pandeyn@iitk.ac.in; chauhan@iitk.ac.in).
	    }\vspace{-7mm}
}

\maketitle 

\begin{abstract}
A simulation-based study of variability of remnant polarization $\left(P_r\right)$ in a multi-granular 3-D ultra-thin ferroelectric (FE) capacitor is presented in this paper. The Poisson Voronoi Tessellation Diagram (PVD) is used for the nucleation of grains in the FE region, which corresponds to the physical growth mechanism. The PVD algorithm implemented in MATLAB is coupled with TCAD simulations, to trace the ferroelectric hysteresis loop.  It is found that the grains which have linear profile of $P_r$ show larger variability in the FE hysteresis loop, compared to the grains, which follow the Gaussian distribution of $P_r$. Additionally, the impact of dielectric content in the FE grains is analyzed. It is seen that the dielectric grains cause very large amount of variability in the FE hysteresis loop. An increase in the dielectric grains also leads to a loss in the retentivity of the hysteresis loop.
\end{abstract}
\section{Introduction}
\label{sec:introduction}
\IEEEPARstart{T}{HE} \label{sec:introduction} 
ferroelectric (FE) materials are being used extensively in the non-volatile memory (NVM) applications \cite{memory_1}-\cite{memory_6} and are also being explored for negative capacitance FET (NCFET) applications \cite{Salahuddin}-\cite{h}. Ferroelectric crystal is most likely to grow in the poly-crystalline (multi-grain) form due to mismatch in lattice constants or defects in the process \cite{grain_1}-\cite{kim} and a fraction of ferroelectric grains always remains in the monoclinic phase, hence, these grains don't exhibit ferroelectricity \cite{Xu},\cite{kao}. The locations of these monoclinic grains introduce an extra source of spatial variability in the remnant polarization ($P_r$) of NCFET \cite{kao}. In order to analyze the variability of remnant polarization with the dielectric (monoclinic) grains in the FE material, a Monte Carlo simulation-based study is proposed in \cite{Lin}. Although
the impact of dielectric grains on the FeFET NVM has been studied in \cite{Liu}, authors in \cite{Liu} and other recent simulation-based studies \cite{kao},\cite{Lin} consider the homogeneous square shape of grains and uniform FE properties in the whole grain, which is not practical. Because, the nucleation of grains in the FE region is completely random and depends on both size and position of the grain \cite{grain_1}-\cite{kim}. 
In this work, we develop a new methodology for the nucleation of grains to model the realistic experimental environment, which takes into account the dependence of nucleation on both the random shape and random position of grains.

\section{Methodology}
\begin{figure}[!t]
	\centering 
	\begin{minipage}{0.5\textwidth}
		\centering
		\includegraphics[width=1.0\textwidth]{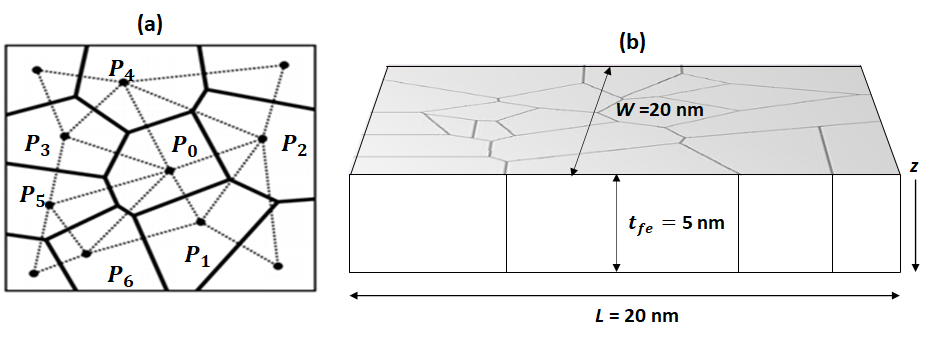}
		\label{fig:prob1_6_1}
	\end{minipage}\vspace{-4mm}
	\caption{(a) Grain distribution from PVD algorithm in a 2-D region where each polygon represents a grain of the FE material. (b) The grains are extruded in the z-direction to form 3-D columnar grains.
	}\label{fig:cap}\vspace{-4mm}
\end{figure}

First, defects points in the FE region are randomly scattered in a 2-D $x-y$ plane. Then Poisson Voronoi (PVD) algorithm  \cite{algo_1},\cite{algo_2} is used to nucleate the grains by these defect points. Fig. \ref{fig:cap}(a) shows the implementation of PVD algorithm in a 2-D bounded region \cite{algo_2}. The points $P_0$..$P_6$ are randomly distributed and these points act as nucleation sites for the grains. Starting from an initial point (e.g. $P_0$), closest neighbor to this point are determined. Subsequently, perpendicular bisectors are drawn on dashed lines (joining lines of neighboring points) to form a closed polygon, which represent a grain. 
\begin{figure}[!t]
	\centering 
	\begin{minipage}{0.7\textwidth}
	\hspace{2mm}
		\includegraphics[width=0.7\textwidth]{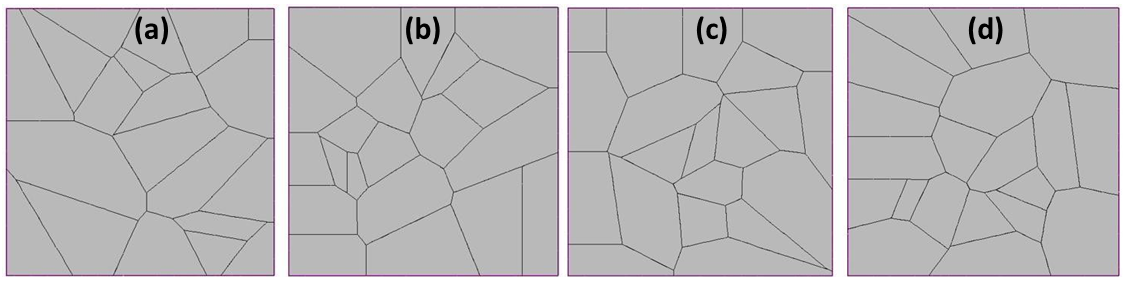}
		\label{fig:prob1_6_1}
	\end{minipage}
	\caption{A 2-D view of grains nucleated by PVD algorithm. The formation of grains in FE is random, hence, each figure shows a different
		grain pattern.
	}\label{fig:grain}\vspace{-7mm}
\end{figure}
Since the size and the position of grains are randomly distributed in the FE region, to study the variability of $P_r$, 75 structures are considered in each simulation (beyond this sample of 75 structures, space replication is observed). 
Fig. \ref{fig:grain} shows an example of the formation of grains in the FE. The nucleated grain is extruded along the z-axis until the FE thickness forms a 3-D grain structure (see Fig. \ref{fig:cap}(b)). Since the ferroelectric thickness is 5 nm, only one grain can be accommodated in the thickness direction \cite{S_kim},\cite{kim}.
\begin{figure}[!t]
	\centering
	\begin{minipage}{0.7\textwidth}
		\centering \hspace{-3.9cm}
		\includegraphics[width=0.75\textwidth]{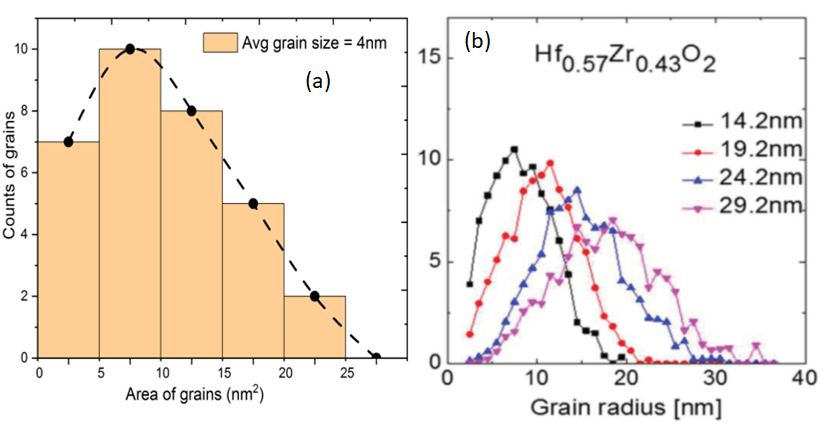}
		\label{fig:prob1_6_1}
	\end{minipage} 
	\caption{(a) Distribution of grains with grain area follows the Gamma distribution. The area of each grain is calculated by MATLAB and the count of the grains (n) are given as, n = 4$\left ( W\times L \right )/\pi\left ( \text{grain size} \right )^2$. (b) Demonstration of a Gamma distribution in the experimentally observed FE grains \cite{grain_2}. 
	}\label{fig:gamma_dis}\vspace{-5mm}
\end{figure}

After the formation of 3-D PVD grains, the $P_r$ is distributed into these grains. The FE grain structures are implemented in MATLAB \cite{matlab}, and these structures are used in 3-D Sentaurus TCAD \cite{tcad}, to capture the electrostatics of the device. A 3 V (peak to peak) and 100 Hz sinusoidal wave is applied at the input terminal of the FE capacitor and Preisach hysteresis model \cite{pre} is used to trace the FE hysteresis loop. The polarization ($P_r$ and $P_s$) are considired to be isotropic in all three coordinates axis. All other parameters in Preisach hysteresis model are taken to be their default values in TCAD Sentaurus \cite{tcad}.
It may be noted here that in the PVD algorithm, each grain is nucleated from a random defect point, which is also observed in the experimental results \cite{grain_1}-\cite{kim}. 

\section{results and discussion}
\subsection{Variability of $P_r$ in Ferroelectric Grains}
Fig. \ref{fig:gamma_dis}(a) shows the plot of number of grains (count of grains) with respect to the area of grains. 
It can be observed in the figure that grains distribution approximately follows the Gamma distribution profile. Fig. \ref{fig:gamma_dis}(b) shows the experimental results from \cite{grain_2}, in which also grain distribution follows a gamma distribution profile, with average grain size of 4.3 nm. Thus, Fig. \ref{fig:gamma_dis}(a) and \ref{fig:gamma_dis}(b) show that the developed PVD algorithm closely matches with the experimental observations.

\begin{figure}[!t]
	\centering 
	\begin{minipage}{0.5\textwidth}
		\centering
		\includegraphics[width=0.65\textwidth]{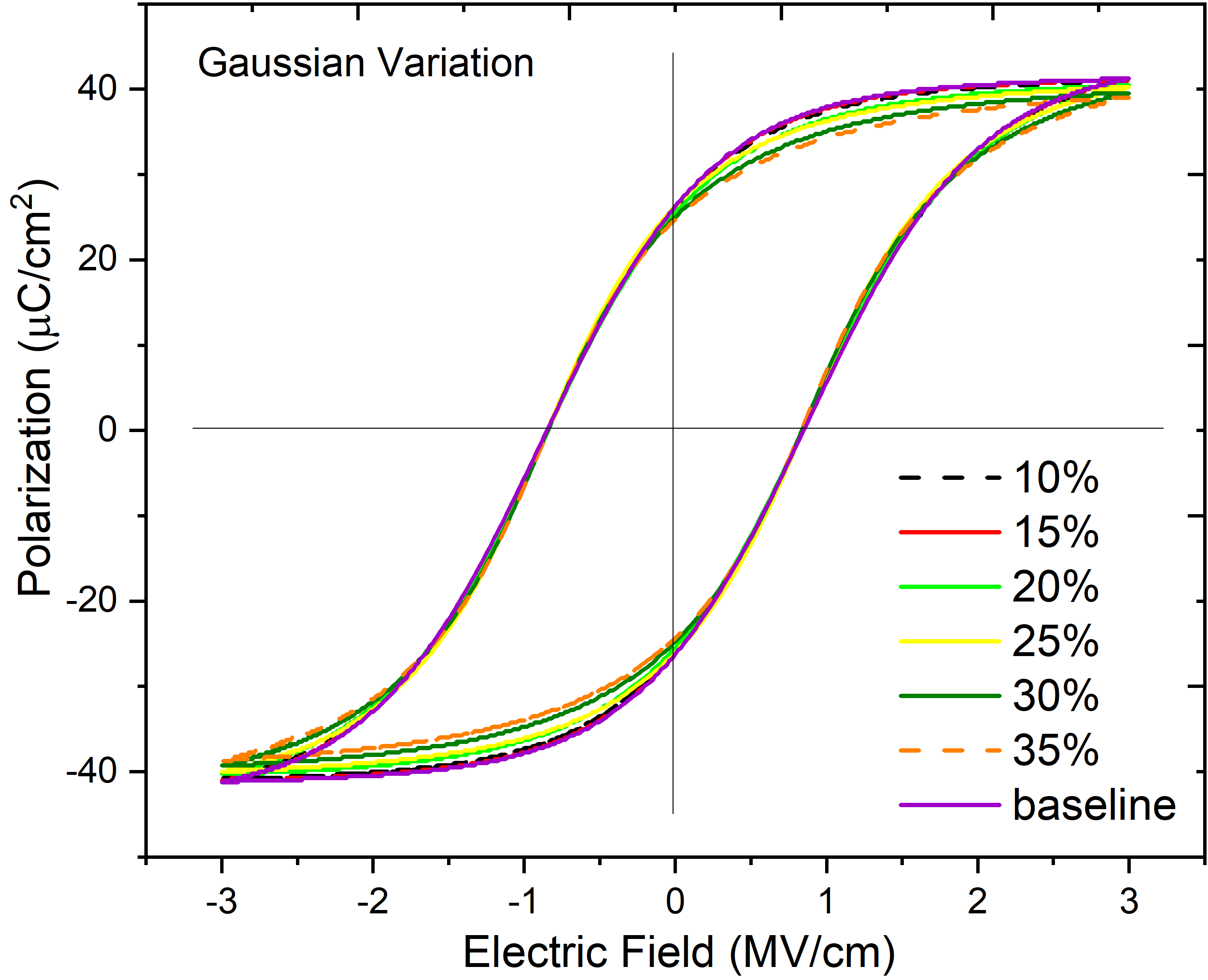}
		\label{fig:prob1_6_1}
	\end{minipage}
	\caption{Variability of $P_r$ in the FE grains. The $P_r$ follows the Gaussian distribution with the grain area. The variability is negligible in P-E loops for a smaller value of variance ($\sigma$). The variability is only observed for $\sigma\geq$ 30\%.  
	}\label{fig:gauss}\vspace{-3mm}
\end{figure}

Experimental data, which shows direct dependence of $P_r$ of Hf$_x$Zr$_{1-x}$O$_2$ (HZO) on the grain area is not available in the literature. However, the properties of FE material of class ABO$_3$ exhibit the size effect \cite{grain_4},\cite{grain_5}. The $P_r$ of these materials varies linearly with grain area. Therefore, to study the variability of $P_r$ of HZO material with respect to the grain area, we consider three different possibilities of $P_r$ distribution: Gaussian, linearly increasing and linearly decreasing.
Fig. \ref{fig:gauss} shows the polarization vs electric field (P-E) characteristic for the Gaussian distribution of $P_r$ with respect to grain area. The median value of $P_{r}$ is 27.67 $\mu$C/cm$^2$ ($P^{m}_{r}$)\cite{hoff}, and the variance ($\sigma$), varies from 0.15$\times P_s$ to 0.35$\times P_s$. Hence, all the grains are allocated a different value of $P_r$ within $\sigma$ range. Total 75 samples of grain structures are considered for each $\sigma$ simulation, and the results are plotted by taking the average of total simulated P-E loops. For smaller $\sigma$, the value of $P_r$ of individual grain lies around the $P^{m}_{r}$. That is why, there is negligible change in P-E curves for small variations in $\sigma$. On the other hand, significant change is observed in the P-E loop, when $\sigma$ approaches to 0.35$P_s$.
\begin{figure}[!t]
	\centering
	\begin{minipage}{0.7\textwidth}
		\centering \hspace{-3cm}
		\includegraphics[width=0.7\textwidth]{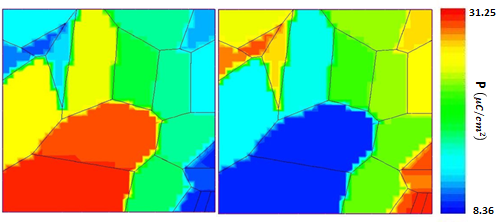}
		\label{fig:prob1_6_1}
	\end{minipage} 
	\caption{2-D surface distribution of $P_r$ in FE grains. In (a) and (b), $P_r$ is increasing and decreasing with the grains area respectively.
	}\label{fig:surf}\vspace{-4mm}
\end{figure}
\begin{figure}[!t]
	\centering 
	\begin{minipage}{0.5\textwidth}
		\centering
		\includegraphics[width=0.65\textwidth]{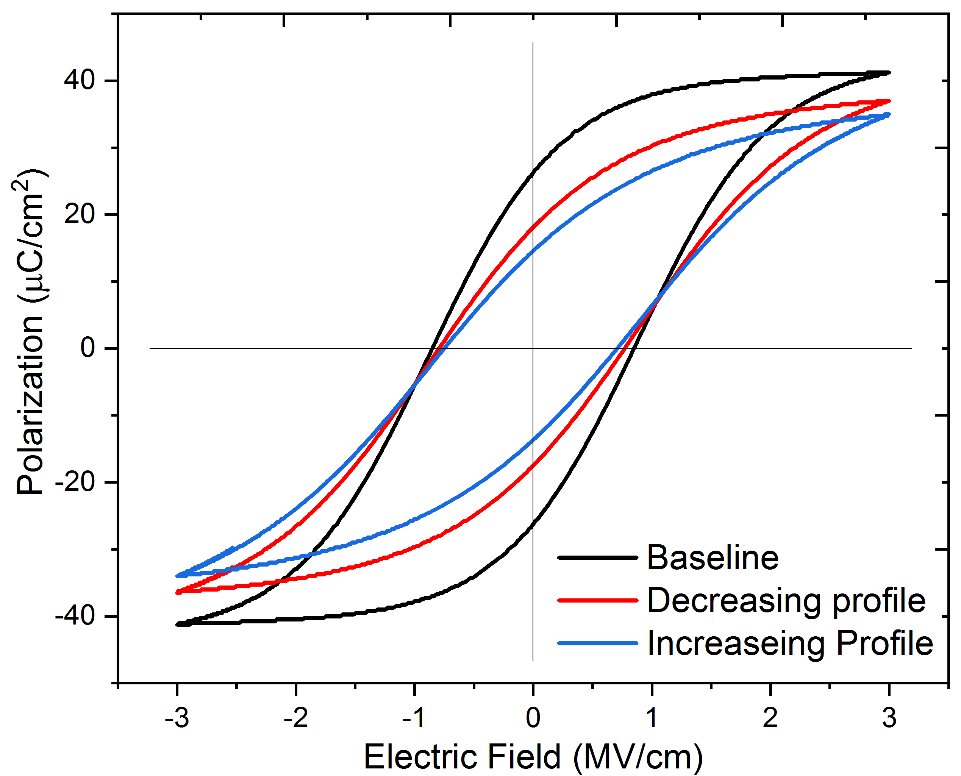}
		\label{fig:prob1_6_1}
	\end{minipage}
	\caption{Variability in the P-E hysteresis loop. The $P_r$ is linearly increasing and decreasing with grain area. The linear dependency of $P_r$ causes a much larger variability in P-E loop than the gaussian profile of $P_r$ with grain area (see Fig. \ref{fig:gauss}). The linear profile of $P_r$ with the grain area is also observed in the experimental results as reported in \cite{grain_4},\cite{grain_5}.
	}\label{fig:lin}\vspace{-4mm}
\end{figure}

Fig. \ref{fig:surf}(a) shows the 2-D surface plot of $P_r$ versus grain area for linearly increasing $P_r$ with grain area. First, we calculate the area of each grain. Then the value of $P_r$, that is proportional to grain area is assigned to individual grain. Fig. \ref{fig:surf}(b) is plotted for linearly decreasing $P_r$ with grain area. The same algorithm is used for all 75 different grain structures. The minimum value of $P_r$ ($P^{min}_{r}$) is $\approx$ 8 $\mu$C/cm$^2$ considered \cite{grain_2}, which corresponds to an undoped HZO (\% Zr = 0) ferroelectric and the maximum value of $P_{r}$ ($P^{max}_{r}$) is 31.25 $\mu$C/cm$^2$ (= 0.75$P_s$). 

\begin{figure}[!t] 
	\begin{minipage}{0.9\textwidth}
		\centering \hspace{-7.4cm}
		\begin{subfigure}[]{0.39\textwidth} 
			\centering
			\includegraphics[width=0.7\textwidth]{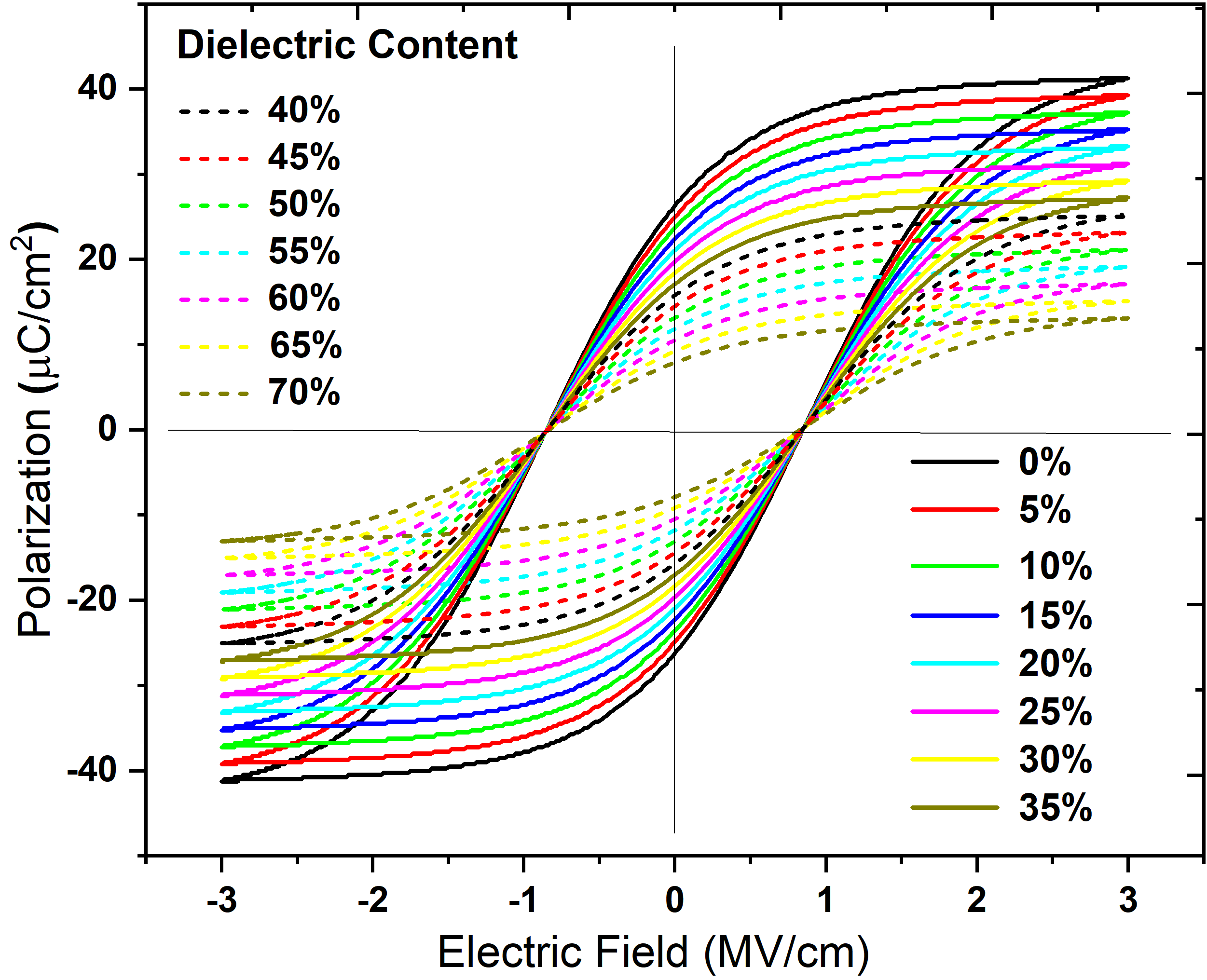}                
			\label{fig:gull}       
		\end{subfigure}
		\hspace{-20mm}
		\begin{subfigure}[]{0.39\textwidth}  
			\centering
			\includegraphics[width=0.7\textwidth]{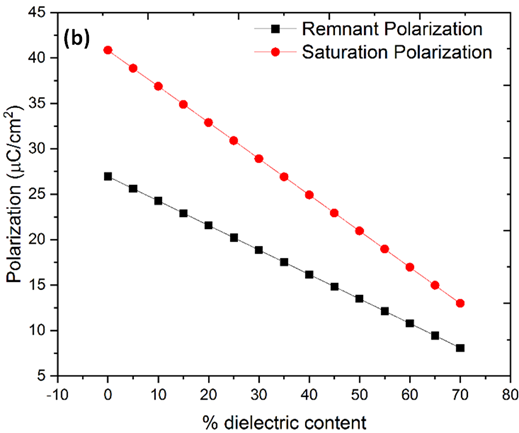}              	\end{subfigure}
	\end{minipage}
	\caption{(a) Impact of dielectric content in P-E hysteresis loop variability. (b) $P_r$ and $P_s$ variations with increasing dielectric content in the FE. As dielectric content increases in the FE material the $P_r$ and $P_s$ decreases which leads to a loss in the memory window of the FE capacitor.} \label{fig:diel}\vspace{-5mm} 
\end{figure}
Fig. \ref{fig:lin} shows the variability in P-E loop for linearly increasing and linearly decreasing $P_r$ with grain area, which is observed experimentally in the ABO$_3$ FE materials \cite{grain_4},\cite{grain_5}. It can be observed in Fig. \ref{fig:lin} that the coercive field ($E_c$) reduces from 0.85 MV/cm to 0.7 MV/cm ($\approx$18\%), which leads to reduction in the hysteresis loop. We can also see that polarization saturates to a lower value than baseline case. Here, the baseline stands for mono-grain FE, which has uniform property in the whole material. Additionally, the value of remnant polarization decreases by 47\% compared to the baseline $P_r$. The red curve in Fig. \ref{fig:lin} corresponds to linearly decreasing $P_r$ with grain area. The values of $P^{max}_{r}$ and $P^{min}_{r}$ are assigned to grains, that have the smallest and largest area respectively (see Fig. \ref{fig:surf}(b)). The variability introduced by linear decreasing $P_r$ (Red curve) is smaller than the linearly increasing $P_r$ (Blue curve). That is why, for linearly decreasing $E_c$, $P_r$ and $P_s$ have values as 0.77 MV/cm, 18 $\mu$C/cm$^2$, and 37 $\mu$C/cm$^2$ respectively (smaller \% change than Blue curve).
\subsection{Dielectric Phase in Ferroelectric Grains}
Some of the grains in the ferroelectric material do not exhibit the ferroelectricity and remain in dielectric phase \cite{Xu}-\cite{Liu}. 
To analyze the impact of dielectric content in FE, a linear relationship of  polarization with the electric field $\left ( P=\epsilon_{0}\left (\epsilon_{r}+1  \right ) E,\text{where, $\epsilon_{r}$ = 22} \right )$ is used in the specific percentage of total number of grains. The assigned maximum percentage to dielectric content is 70\% of the total grains and the assigned minimum percentage is zero (pure FE material without any dielectric grains). In this work, we are simulating an isolated FE capacitor, hence, the positions of dielectric grains do not introduce any extra source of variability. 
Fig. \ref{fig:diel}(a) shows that as the dielectric content in the FE material increases, the P-E loops start to deteriorate and, hence, the ferroelectricity of the device starts to decreases. Fig. \ref{fig:diel}(b) shows the variations in $P_r$ and $P_s$ with dielectric content. Both decrease linearly with increment in the \% of dielectric content in FE. The value of $P_r$ approaches to that of baseline FE with \% of Zr = 0, \cite{grain_2}) for dielectric content equal to the 60\% in ferroelectric of total number of grains. 

\section{conclusion}
The PVD algorithm is used to nucleate grains at randomly scattered defects points. Size and positions of nucleated grains are considered to be random in simulations. Hence, simulated results are remarkably close to the experimental observations. We have shown that the variability in the P-E hysteresis loop is small for the Gaussian distribution of $P_r$ with respect to grain area, due to small amount of variations in $P_r$ of adjacent grains. However, variability in P-E loop increases significantly, when $P_r$ follows the linear profile with respect to grain area. Additionally, both $P_s$ decreases and hysteresis width reduces, hence, ferroelectricity (memory window) reduces. Therefore, experimentally fabricated FE device (which has variability of $P_r$ in grains) will show poor retentivity than the mono-grain (uniform property in the whole material) FE device. Further, it is found that, the retentivity of FE memory decreases, as dielectric content in FE material increases.

\end{document}